\documentclass[final,runningheads,a4paper,orivec,UKenglish,envcountsame,envcountsect]{llncs}
\pdfoutput=1

\usepackage[T1]{fontenc}
\usepackage[utf8]{inputenc}
\usepackage{bbding}

\makeatletter
\renewcommand\section{\@startsection{section}{1}{\z@}%
                       {-12\p@ \@plus -4\p@ \@minus -4\p@}%
                       {6\p@ \@plus 4\p@ \@minus 4\p@}%
                       {\normalfont\large\bfseries\boldmath
                        \rightskip=\z@ \@plus 8em\pretolerance=10000}}

\renewcommand\subsubsection{\@startsection{subsubsection}{3}{\z@}%
                       {-6\p@ \@plus -2\p@ \@minus -2\p@}%
                       {-0.5em \@plus -0.22em \@minus -0.1em}%
                       {\normalfont\normalsize\bfseries\boldmath}}
\makeatother

\usepackage[paperheight=235mm,paperwidth=155mm,textwidth=12.2cm,textheight=19.3cm]{geometry}

\usepackage{ifdraft}
\ifdraft{
\usepackage[marginparwidth=2cm,nohead,nofoot,
            headsep = 1cm,
            footskip = 1cm,
            textwidth=12.2cm,
            textheight=19.3cm,
            paperwidth=155mm
            paperheight=235mm
            marginparsep=1mm,
            marginparwidth=1.8cm,
            footskip=1cm,
            centering
            ]{geometry}
          }{}

\usepackage[inline]{enumitem}
\setlist[enumerate,1]{label=(\arabic*),font=\normalfont,align=left,leftmargin=0pt,labelindent=0pt,listparindent=\parindent,labelwidth=0pt,itemindent=!,topsep=0pt,parsep=0pt,itemsep=0pt,start=1}
\setlist[enumerate,2]{label=(\alph*),font=\normalfont,labelindent=*,leftmargin=*,start=1}
\setlist[itemize]{labelindent=*,leftmargin=*,topsep=5pt,itemsep=3pt}
\setlist[description]{labelindent=*,leftmargin=*,itemindent=-1 em}

\usepackage[%
rm={oldstyle=false,proportional=true},%
sf={oldstyle=false,proportional=true},%
tt={oldstyle=false,proportional=true,variable=true},%
qt=false%
]{cfr-lm}

\usepackage{microtype}

\usepackage[noadjust]{cite}

\usepackage{seqsplit}
\usepackage{xstring}

\makeatletter
\usepackage[notcite,notref]{showkeys}%

\newcommand\noshowkeys{\def\hideNextShowKeysLabel{test}}
\renewcommand*\showkeyslabelformat[1]{%
\@ifundefined{hideNextShowKeysLabel}{%
\noexpandarg%
\StrSubstitute{#1}{ }{\textvisiblespace}[\TEMP]%
\parbox[t]{\marginparwidth}{\raggedright\normalfont\small\ttfamily\(\{\){\color{red!50!black}\expandafter\seqsplit\expandafter{\TEMP}}\(\}\)}%
}{}%
}
\makeatother

\makeatletter
\RequirePackage[bookmarks,unicode,colorlinks=true]{hyperref}%
   \def\@citecolor{blue}%
   \def\@urlcolor{blue}%
   \def\@linkcolor{blue}%

\def\orcidID#1{\smash{\href{http://orcid.org/#1}{\protect\raisebox{-1.25pt}{\protect\includegraphics{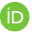}}}}}
\makeatother

\hypersetup{
  pdftitle={Distributed Coalgebraic Partition Refinement},
  pdfauthor={Fabian Birkmann, Hans-Peter Deifel, Stefan Milius},
  pdfduplex={DuplexFlipLongEdge},
}

\ifdraft{%
  \makeatletter
  \setcounter{tocdepth}{2}
  \makeatother
}{}

\usepackage{etoolbox} %
\makeatletter
\newenvironment{notheorembrackets}{%
\csdef{@spopargbegintheorem}##1##2##3##4##5{\trivlist%
      \item[\hskip\labelsep{##4##1\ ##2}]{##4{##3}\@thmcounterend\ }##5}%
    }{%
\csdef{@spopargbegintheorem}##1##2##3##4##5{\trivlist%
      \item[\hskip\labelsep{##4##1\ ##2}]{##4(##3)\@thmcounterend\ }##5}%
    }
\makeatother

\newcommand{\takeout}[1]{\empty}

\usepackage{amsmath}
\usepackage{amssymb}
\usepackage[final]{listings}
\usepackage{multirow}
\usepackage{pgfplots}
\usepackage{wrapfig}
\pgfplotsset{compat=1.5}
\usepackage{tikz}
\usepackage[linesnumbered]{algorithm2e}
\newcounter{lastAlgoLine}
\usepackage{algorithmic}
\usepackage{dsfont}
\usepackage{xspace}
\usepackage{subcaption}
\usepackage{booktabs}
\usepackage{tikz}
\usetikzlibrary{cd, calc, shapes, automata, fit, decorations.pathreplacing}

\tikzset{
  coalgebra drawing/.style={
    state/.append style={
      minimum width=0pt,
      minimum height=0pt,
      inner sep=0.8mm,
    },
    every edge/.append style={
      shorten <= 1pt,
      shorten >= 1pt,
    }
  }
}

\tikzset{shiftarr/.style={
        rounded corners,%
        to path={--([#1]\tikztostart.center)
                     -- ([#1]\tikztotarget.center) \tikztonodes
                     -- (\tikztotarget)},
}}

\usepackage{fixme}
\fxuselayouts{footnote,marginclue,nomargin}
\fxusetheme{color}
\FXRegisterAuthor{fb}{afb}{FB}
\FXRegisterAuthor{sm}{asm}{SM}
\FXRegisterAuthor{hpd}{hpd}{HPD}

\newcommand{\copar}{\textsf{CoPaR}\xspace}

\SetKwInOut{Variables}{Variables}
\SetKwInOut{Initialise}{Initialise}
\SetKwInOut{Finalise}{Finalise}

\newcommand{\hash}{\mathsf{hash}}  %
\newcommand{\sig}[1]{\mathsf{sig}_{#1}}
\newcommand{\ID}{\textsf{ID}\xspace}

\newcommand{\inlist}{\mathsf{In}_{s}}
\newcommand{\inlistof}[1]{\mathsf{In}_{#1}}

\newcommand{\R}{\mathds{R}}
\newcommand{\N}{\mathds{N}}
\newcommand{\Z}{\mathds{Z}}
\newcommand{\xra}[1]{\xrightarrow{~#1~}}
\newcommand{\fpair}[1]{\langle#1\rangle}
\newcommand{\id}{\mathsf{id}}

\newcommand{\Set}{\ensuremath{\mathsf{Set}}\xspace}
\newcommand{\Pow}{{\mathcal{P}_\omega}}
\newcommand{\Bag}{\mathcal{B}}
\newcommand{\Dist}{\ensuremath{{\mathcal{D}_\omega}}\xspace}
\newcommand{\Monval}[1]{M^{(#1)}}

\newcommand{\set}[1]{\{#1\}}
\newcommand{\op}[1]{\ensuremath{\mathsf{#1}}}
\newcommand{\inj}{\op{in}}
\newcommand{\pr}{\op{pr}}
\newcommand{\sigop}{\op{sig}}

\newcommand{\CO}{\mathcal{O}}

\spnewtheorem{thm}[theorem]{Theorem}{\bfseries}{\itshape}
\spnewtheorem{cor}[theorem]{Corollary}{\bfseries}{\itshape}
\spnewtheorem{cnj}[theorem]{Conjecture}{\bfseries}{\itshape}
\spnewtheorem{lem}[theorem]{Lemma}{\bfseries}{\itshape}
\spnewtheorem{lemdefn}[theorem]{Lemma and Definition}{\bfseries}{\itshape}
\spnewtheorem{prop}[theorem]{Proposition}{\bfseries}{\itshape}
\spnewtheorem{defn}[theorem]{Definition}{\bfseries}{\upshape}
\spnewtheorem{rem}[theorem]{Remark}{\bfseries}{\upshape}
\spnewtheorem{notation}[theorem]{Notation}{\bfseries}{\upshape}
\spnewtheorem{expl}[theorem]{Example}{\bfseries}{\upshape}
\spnewtheorem{thmdefn}[theorem]{Theorem and Definition}{\bfseries}{\itshape}
\spnewtheorem{propdefn}[theorem]{Proposition and Definition}{\bfseries}{\itshape}
\spnewtheorem{assumption}[theorem]{Assumption}{\bfseries}{\upshape}
\spnewtheorem{algo}[theorem]{Algorithm}{\bfseries}{\upshape}
\spnewtheorem{constr}[theorem]{Construction}{\bfseries}{\upshape}

 \renewenvironment{definition}{\begin{defn}}{\end{defn}}
 \renewenvironment{remark}{\begin{rem}}{\end{rem}}
 \renewenvironment{example}{\begin{expl}}{\end{expl}}

\definecolor{FAU-Blau}{RGB}{0,47,108}					%
\definecolor{FAU-Dunkelblau}{RGB}{4,30,66}				%
\definecolor{FAU-Phil-Gelb}{RGB}{255, 184,28}			%
\definecolor{FAU-Phil-Orange}{RGB}{232,119,34}			%
\definecolor{FAU-RW-Rot}{RGB}{200,16,46}				%
\definecolor{FAU-RW-Dunkelrot}{RGB}{151,27,47}			%
\definecolor{FAU-Med-Blau}{RGB}{0,163,224}				%
\definecolor{FAU-Med-Dunkelblau}{RGB}{0,97,160}			%
\definecolor{FAU-Nat-Gruen}{RGB}{67,176,42}				%
\definecolor{FAU-Nat-Dunkelgruen}{RGB}{34,136,72}		%
\definecolor{FAU-Tech-Metallic}{RGB}{119,159,181}		%
\definecolor{FAU-Tech-Dunkelmetallic}{RGB}{65,116,141}	%

\title{Distributed Coalgebraic Partition Refinement}
\titlerunning{Distributed Coalgebraic Partition Refinement}
\author{Fabian Birkmann\orcidID{0000-0001-5890-9485}%
  \and
  Hans-Peter Deifel%
  \Envelope\fnmsep{}\thanks{Supported by the Deutsche Forschungsgemeinschaft (DFG) within the Research and Training Group 2475 ``Cybercrime and  Forensic Computing'' (393541319/GRK2475/1-2019)}%
  \orcidID{0000-0002-9542-9664}
  \and
  Stefan Milius%
  \thanks{Supported by Deutsche Forschungsgemeinschaft (DFG) under project MI~717/7-1.}\orcidID{0000-0002-2021-1644}
}
\authorrunning{F.~Birkmann, H.-P.~Deifel, S.~Milius}

\institute{Friedrich-Alexander-Universität Erlangen-Nürnberg, Germany
\email{\{fabian.birkmann,hans-peter.deifel,stefan.milius\}@fau.de}
}

\begin{document}
\maketitle

\begin{abstract}
  Partition refinement is a method for minimizing automata and transition
  systems of various types. Recently, a new partition refinement
  algorithm and associated tool \copar were developed that are generic in the transition type of the
  input system and match the theoretical run time of the best known algorithms
  for many concrete system types. Genericity is achieved by modelling transition
  types as functors on sets and systems as coalgebras. Experimentation has shown
  that memory consumption is a bottleneck for handling systems with a large
  state space, while running times are fast. We have therefore extended an
  algorithm due to Blom and Orzan, which is suitable for a distributed
  implementation to the coalgebraic level of genericity, and implemented it in
  \copar. Experiments show that this allows to handle much larger state spaces.
  Running times are low in most experiments, but there is a significant penalty
  for some.
\end{abstract}

\section{Introduction}
\label{sec:intro}

Minimization is an important and basic algorithmic task on state-based
systems, concerned with reducing the state space as much as possible
while retaining the system's behaviour. It is used for
equivalence checking of systems and as a subtask in model checking
tools in order to handle larger state spaces and thus mitigate the
state-explosion problem.

We focus on the task of identifying behaviourally equivalent states
modulo bisimilarity. For classic labelled transitions systems this
notion obeys the principle `states $s$ and $t$ are bisimilar if for
every transition $s \xra{a} s'$, there exists a transition
$t \xra{a} t'$ with $s'$ and $t'$ bisimilar', and symmetrically for
transitions from $t$. Bisimilarity is a rather fine-grained
branching-time notion of equivalence~(cf.~\cite{vGlabbeek01}); it is
widely used and preserves all properties expressible as $\mu$-calculus
formulas.%
\smnote{}  Moreover, it has been generalized to
yield equivalence notions for many other types of state-based systems
and automata.

Due to the above principle, bisimilarity is defined by a fixed point,
to be understood as a greatest fixed point and is hence approximable from above.
This is used by
\emph{partition refinement} algorithms:
The initial partition considers all states tentatively equivalent is then
iteratively  refined using observations about the states until a
fixed point is reached.
Consequently, such procedures run in
polynomial time and can also be efficiently implemented, in contrast
to coarser system equivalences such as trace equivalence and language
equivalence of nondeterministic systems which are
PSPACE-complete~\cite{KanellakisS90}. This makes minimization under
bisimilarity interesting even in cases where the main equivalence is
linear-time, such as for automata.

Efficient partition refinement algorithms exist for various systems:
Kanellakis and Smolka provide a minimization algorithm with run time
$\CO(m \cdot n)$ for labelled transition systems with $n$ states and
$m$ transitions. Even faster algorithms have been developed over the
past 50 years for many types of systems. For example, Hopcroft's
algorithm for minimizing deterministic automata has run time in
$\CO(n \cdot \log n)$~\cite{Hopcroft71}; it was later generalized to
variable input alphabets, with run time $\CO(n\cdot |A|\cdot \log n)$
\cite{Gries1973,Knuutila2001}. The Paige-Tarjan algorithm minimizes
transition systems in time $\CO((m+n)\cdot \log n)$
\cite{PaigeTarjan87}, and generalizations to labelled transition
systems have the same time complexity
\cite{HuynhTian92,DerisaviEA03,Valmari09}. For the minimization of
weighted systems (a.k.a.~\emph{lumping}), Valmari and
Franchescini~\cite{ValmariF10} have developed a simple
$\CO((m+n)\cdot \log n)$ algorithm for systems with rational
weights. Buchholz~\cite{Buchholz08} gave an algorithm for weighted
automata, and Högberg et al.~\cite{HoegbergEA07} one for (bottom-up)
weighted trees automata, both with run time in $\CO(m \cdot n)$. 

\enlargethispage{5pt}
In previous work~\cite{dmsw17,wdms20}, an efficient
partition refinement algorithm was provided which is generic in the system type,
captures all the above system types, and matches or, in some cases
even improves on the run time complexity of the respective
specialized algorithms. Subsequently, we have shown how to extend the
generic complexity analysis to weighted tree automata and
implemented the algorithm in the tool \copar~\cite{DeifelEA19,wdms21},
again matching the previous best run time complexity and improving it
in the case of weighted tree automata with weights from a
non-cancellative monoid. The algorithm is based on ideas of Paige
and Tarjan, which leads to its efficiency. Genericity is achieved by
modelling state based systems as coalgebras, following the paradigm of
universal coalgebra~\cite{Rutten00}, in which the transitions
structure of systems is encapsulated by a set functor. The algorithm
and tool are \emph{modular} in the sense that functors can be built
from a preimplemented set of basic functors by standard set
constructions such as cartesian product, disjoint union and functor
composition. The tool then automatically derives a parser for input
coalgebras of the composed type and provides a corresponding partition
refinement implementation off the shelf. In addition, new basic
functors $F$ may easily be added to the set of basic functors by
implementing a simple refinement interface for them plus a parser for
encoded~$F$-coalgebras. Our experiments with the tool have shown that
run time scales well with the size of systems. However, memory usage
becomes a bottleneck with growing system size, a problem that has
previously also been observed by Valmari~\cite{Valmari10SimpleBisim}
for partition refinement. One strategy to address this is to
distribute the algorithm across multiple computers, which store and
process only a part of the state space and communicate via message
passing. For ordinary labelled transition systems and Markov systems
this has been investigated in a series of
papers by Blom and Orzan~\cite{BlomO02,BlomO03,BlomO03b,bo05-01,bo05-06,bhkp08} who were
also motivated to mitigate the memory bottleneck of sequential
partition refinement algorithms.

Our contribution in this paper is an extension of \copar by an
efficient distributed partition algorithm in coalgebraic
generality. Like in Blom and Orzan's work, our algorithm is a
distributed version of a simple but effective algorithm called ``the
naive method''~\cite{KanellakisS90}, or ``the final chain algorithm''
in coalgebraic generality~\cite{KonigK14,wdms20}. We first generalize
signature refinement introduced by Blom and Orzan to the level of
coalgebras. We also combine generalized signatures
(\autoref{sec:coalgpartref}) with the previous encodings of set
functors and their coalgebras~\mbox{\cite{DeifelEA19,wdms21}} via the
new notion of a signature interface (\autoref{D:sig-int}). This is a
key idea to make coalgebraic signature refinement and the final chain
algorithm implementable in a tool like \copar. In addition, we
demonstrate how signature interfaces of functors can be combined
(\autoref{C:comb-sig} and \autoref{P:comb-sig}) along standard functor
constructions. This yields a similar modularity principle than for the
previous sequential algorithm. However, this is a new feature for
signature refinement and also, to our knowledge, for the final chain
algorithm. Consequently, our distributed, modular and generic implementation of the
final chain algorithm is new (already as sequential algorithm).

We also provide experiments demonstrating its scalability and show
that much larger state spaces can indeed be handled. Our benchmarks
include weighted tree automata for non-cancellative monoids, a type of
system for which %
our previous sequential implementation is
heavily limited by its memory requirements. For those systems the running times
of the distributed algorithm are even faster then those of the
sequential algorithm. In a second set of benchmarks stemming from the
PRISM benchmark suite~\cite{KNP12} we again show that larger systems
can now be handled; however, for some of these there is a penalty in
run time.

\smnote[inline]{}

\subsubsection{Related work.} Balcazar et al.~\cite{BalcazarEA92} have
proved that the problem of bisimilarity checking for labelled transition
systems is $P$-complete, which implies that it is hard to
parallelize efficiently. Nevertheless, parallel algorithms have been
proposed by Rajasekaran and Lee~\cite{RajasekaranL98}. These are
designed for shared memory machines and hence do not distribute RAM
requirements over multiple machines.

Symbolic techniques are an orthogonal approach to reduce memory
usage of partition refinement algorithms and have been explored e.g.~by
Wimmer~et~al.~\cite{Wimmer2006} and van~Dijk and
de~Pol~\cite{vandDjk2018}.

\takeout{}%

Two other orthogonal extensions of the generic coalgebraic
minimization and \copar have been presented in recent work.
First a non-trivial extension computes (1)~reachable
states and (2)~the transition structure of the minimized
systems~\cite{dmw21}. Second, Wißmann et al.~\cite{wms21} have shown
how to compute distinguishing formulas in a Hennessy-Milner style
logic for a pair of behaviourally inequivalent states. 
\takeout{}%

\section{Preliminaries}
\label{sec:prelim}

Our algorithmic framework and the tool \copar~\cite{wdms20,wdms21} are
based on modelling state-based systems abstractly as \emph{coalgebras}
for a (set) \emph{functor} that encapsulates the transition type,
following the paradigm of \emph{universal
  coalgebra}~\cite{Rutten00}. We now recall some standard notations
for sets and maps and basic notions and examples in coalgebra. We fix
a singleton set $1=\{*\}$; for every set~$X$ we have a unique map
$!\colon X\to 1$ and the identity map $\id_X\colon X\to X$. We denote
composition of maps by $(-)\cdot(-)$, in applicative order. Given maps
$f\colon X\to A$, $g\colon X\to B$ we define
$\fpair{f,g}\colon X\to A\times B$ by $\fpair{f,g}(x) =
(f(x),g(x))$. The type of transitions of states in a system is modelled
by a set functor $F$. Informally, $F$ assigns to every set $X$ a set
$FX$ of structured collections of elements of $X$, and an
$F$-coalgebra is a map $c\colon S\to FS$ which assigns to every state $s \in S$ in a
system a structured collection $c(s) \in FS$ of successor states of
$s$. The functor $F$ also determines a canonical notion of behavioural
equivalence of states of a coalgebra; this arises by stipulating that
morphisms of coalgebras are behaviour preserving maps.
\begin{definition}
  A \emph{functor} $F\colon \Set\to\Set$ assigns to each
  set~$X$ a set~$FX$ and to each map $f\colon X\to Y$ a map
  $Ff\colon FX\to FY$, preserving identities and composition
  ($F\id_X=\id_{FX}$, $F(g\cdot f)=Fg\cdot Ff$).  An
  \emph{$F$-coalgebra} $(S,c)$ consists of a set~$S$ of
  \emph{states} and a \emph{transition structure} $c\colon S\to FS$.
  A \emph{morphism} $h\colon (S,c)\to (S',c')$ of $F$-coalgebras is a
  map $h\colon S\to S'$ that preserves the transition structure, i.e.~$Fh\cdot c
  = c'\cdot h$. Two states $s,t\in S$ of a coalgebra $c\colon S\to FS$ are
  \emph{behaviourally equivalent} ($s \sim t$) if there exists a coalgebra
  morphism $h$ with $h(s) = h(t)$.
\end{definition}
\begin{example}\label{E:func}
  We mention several types of systems which are instances of the
  general notion of coalgebra and the ensuing notion of behavioural
  equivalence. All these are possible input systems for our tool
  \copar.%
  \begin{enumerate}
  \item \label{E:func:1} Transition systems. The \emph{finite
      powerset} functor $\Pow$ maps a set~$X$ to the set~$\Pow X$ of
    all \emph{finite} subsets of~$X$, and a map $f\colon X\to Y$ to
    the map $\Pow f = f[-]\colon \Pow X\to \Pow Y$ taking direct
    images.  Coalgebras for $\Pow$ are finitely branching (unlabelled)
    transition systems. Two states are behaviourally equivalent iff
    they are (strongly) bisimilar in the sense of
    Milner~\cite{Milner80,Milner89} and Park~\cite{Park81}. Similarly,
    finitely branching labelled transition systems with label alphabet
    $A$ are coalgebras for the functor $FX = \Pow(A \times X)$.

  \item Deterministic automata. For an input alphabet $A$, the functor given by
    $FX=2\times X^A$, where $2 = \{0,1\}$, sends a set $X$ to the set
    of pairs of boolean values and functions $A\to X$. An
    $F$-coalgebra $(S,c)$ is a deterministic automaton (without an
    initial state). For each state $s\in S$, the first component of
    $c(s)$ determines whether $s$ is a final state, and the second
    component is the successor function $A\to S$ mapping each input
    letter $a\in A$ to the successor state of $s$ under input letter
    $a$. States $s,t \in S$ are behaviourally equivalent iff they
    accept the same language in the usual sense.

  \item Weighted tree automata simultaneously generalize tree automata
    and weight\-ed (word) automata. Inputs of such automata stem from a
    finite \emph{signature} $\Sigma$, i.e.~a finite set of input
    symbols, each with a prescribed natural number, its
    \emph{arity}. \emph{Weights} are taken from a commutative monoid
    $(M,+,0)$. A (bottom-up) \emph{weighted tree automaton} (WTA)
    (over $M$ with inputs from $\Sigma$) consists of a finite set $S$
    of states, an output map $f\colon S\to M$, and for each $k\ge 0$,
    a transition map $\mu_k\colon \Sigma_k\to M^{S^k\times S}$, where
    $\Sigma_k$ denotes the set of $k$-ary input symbols in $\Sigma$;
    the maximum arity of symbols in~$\Sigma$ is called the
    \emph{rank}.

    Every signature $\Sigma$ gives rise to its associated
    \emph{polynomial functor}, also denoted~$\Sigma$, which assigns to
    a set $X$ the set $\coprod_{n \in \N} \Sigma_n \times X^n$, where
    $\coprod$ denotes disjoint union (coproduct). Further, for a given
    monoid $(M,+,0)$ the \emph{monoid-valued functor}
    $\Monval -$ sends a set $X$ to the set of maps
    $f\colon X\to M$ that are finitely supported, i.e.~$f(x) = 0$ for
    almost all $x\in X$. Given a map $f\colon X\to Y$, 
    $\Monval f\colon \Monval X\to \Monval Y$ sends a map
    $v\colon X\to M$ in $\Monval X$ to the map
    $y\mapsto \sum_{x\in X, f(x) = y} v(x)$, corresponding to the
    standard image measure construction.

    Weighted tree automata are coalgebras for
    the composite functor $FX = M \times \Monval{\Sigma X}$; indeed, given a
    coalgebra $ c = \fpair{c_1,c_2}\colon S \to M \times
    \Monval{\Sigma S}$, its first component $c_1$ is the output map,
    and the second component $c_2$ is equivalent to the
    family of transitions maps $\mu_k$ described above.

    As proven by Wißmann et al.~\cite[Prop.~6.6]{wdms21}, the
    coalgebraic behavioural equivalence is precisely backward
    bisimulation of weighted tree automata as introduced by Högberg et
    al.~\cite[Def.~16]{HoegbergEA07}.

  \item The \emph{bag functor} $\Bag\colon \Set \to \Set$ sends a set $X$
    to the set of all finite multisets (or \emph{bags}) over $X$. This
    is the special case of the monoid-valued functor for the monoid
    $(\N, + , 0)$. Accordingly, $\Bag$-coalgebras are weighted
    transition systems with positive integers as weights, or they may
    be regarded as finitely branching transition systems where
    multiple transitions between a pair of states are
    allowed. Behavioural equivalence coincides with 
    weighted (or strong) bisimilarity.%
    \smnote{}

  \item Markov chains. The \emph{finite distribution functor} $\Dist$
    is a subfunctor of the monoid-valued functor $\R^{(-)}$ for the
    usual monoid of addition on the real numbers. It maps a set $X$ to
    the set of all finite probability distributions on~$X$. That means
    that $\Dist X$ is the set of all finitely supported maps
    $d\colon X \to [0,1]$ such that $\sum_{x \in X} d(x) = 1$. The
    action of $\Dist$ on maps is the same as that of~$\R^{(-)}$.

    As shown by Rutten and de~Vink~\cite{RuttenDV99}, coalgebras
    $c\colon S \to (\Dist S + 1)^A$ are precisely Larsen and
    Skou's probabilistic transition systems~\cite{LarsenS91}
    (aka.~labelled Markov chains~\cite{DesharnaisEA02}) with the label
    alphabet $A$. In fact, for each state $s \in S$ and action label
    $a \in A$, that state either cannot perform an $a$-action
    (when $c(s)(a) \in 1$) or the distribution $c(s)(a)$
    determines for every state $t \in C$ the probability with which
    $s$ transitions to $t$ with an $a$-action.

    Coalgebraic behavioural equivalence is precisely probabilistic
    bisimilarity in the sense of Larsen and Skou, see 
    Rutten and de Vink~\cite[Cor.~4.7]{RuttenDV99}.

  \item Markov decision processes are systems which feature both
    non-deterministic and probabilistic branching. They are
    coalgebras for composite functors such as
    $\Pow(A \times \Dist(-))$ or
    $\Pow(\Dist(A \times (-))$ (simple/general Segala systems); Bartels
    et al.~\cite{BartelsEA03} list further functors for various species of probabilistic systems.
  \end{enumerate}
\end{example}
\vspace*{-10pt}
\subsubsection{Encodings.}\label{sec:encodings}
To supply coalgebras as inputs to \copar and in order to speak about
the size of a coalgebra in terms of states and transitions, we need
\begin{notheorembrackets}
\begin{definition}[{\cite[Def.~3.1]{dmw21}}]\label{D:encoding}
  An \emph{encoding} of a set functor $F$ consists of a set~$A$ of \emph{labels} and a
  family of maps $\flat_X\colon FX\to\Bag(A\times X)$, one for every set $X$,
  such that the map
  $\fpair{F!, \flat_X}\colon FX \to F1\times\Bag(A\times X)$
  is injective.

  The \emph{encoding} of a coalgebra $c\colon S \to FS$ %
  is $\fpair{F!, \flat_S}\cdot c\colon S\to F1\times \Bag(A\times
  S)$. For $s \in S$ we write $s \xra{a} t$ whenever $(a,t)$ is
  contained in the bag $\flat_S(c(s))$. The \emph{number of states}
  and \emph{edges} of a given encoded input coalgebra are $n = |S|$
  and $m = \sum_{s \in S} |\flat_S(c(s))|$, respectively,
  where $|b| = \sum_{x \in X} b(x)$ for a bag $b\colon X \to \N$.
\end{definition}
\end{notheorembrackets}
An encoding of a set functor $F$ specifies how $F$-coalgebras are
represented as directed graphs, and the required injectivity ensures
that different coalgebras have different encodings.

\begin{example}
  We recall a few key examples of encodings used by
  \copar~\cite{wdms20}; for the required injectivity,
  see~\cite[Prop.~3.3]{dmw21}.
  \begin{enumerate}
  \item For the finite powerset functor $\Pow$ one takes a singleton
    label set $A= 1$ and $\flat_X\colon \Pow X \to \Bag(1 \times X)$
    is the obvious inclusion: $\flat_X(U)(*,x) = 1$ iff
    $x \in U \subseteq X$.
  \item For the monoid-valued functor $\Monval -$ we take labels $A =
    M$, and the map $\flat_X\colon \Monval X \to \Bag(M \times X)$ is given
    by $\flat_X(t)(m,x) = 1$ if $t(x) = m \neq 0$ and $0$ else.
  \item As a special case, the bag functor $\Bag$ has labels $A =
    \N$, and the map $\flat_X\colon \Bag X \to \Bag(\N \times X)$ is given by
    $\flat_X(t)(n,x) = 1$ if $t(x) = n$ and $0$ else.%
    \smnote{}
  \end{enumerate}
\end{example}
\begin{rem}
  \begin{enumerate}
  \item Readers familiar with category theory may wonder about the
    \emph{naturality} of encodings $\flat_X$. It turns
    out~\cite{dmw21} that in almost all instances, our encodings are
    not natural transformations, except for polynomial functors. As
    shown in \emph{op.~cit.}, all our encodings satisfy a property
    called \emph{uniformity}, which implies that they are subnatural
    transformations~\cite[Prop.~3.15]{dmw21}.
    
  \item Having an encoding of a set functor $F$ does not imply a
    reduction of the problem of minimizing $F$-coalgebras to
    that of coalgebras for $\Bag(A \times -)$. In fact, the 
    behavioural equivalence of $F$-coalgebras and coalgebras for
    $\Bag(A \times -)$ may be very different unless $\flat_X$ is
    natural, which is not the case for most encodings. 
  \end{enumerate}
\end{rem}
Functors in \copar can be combined by product, coproduct or composition,
leading to modularity. But in order to automatically handle combined
functors, our tool crucially depends on the ability to form products and
coproducts of encodings~\cite{wdms20,wdms21}. We refrain from going into
technical details, but note for further use that given a pair of
functors~$F_1,F_2$ with encodings $A_i, \flat_{X,i}$ one obtains
encodings for the functors $F_1 \times F_2$ (cartesian product) and
$F_1 + F_2$ (disjoint union) with the label set $A = A_1  + A_2$.
\subsubsection{Input syntax and processing.}\label{sec:input}
We briefly recall the
input format of \copar and how inputs are processed; for more details
see~\cite[Sec.~3.1]{wdms21}. \copar accepts input files representing a
finite $F$-coalgebra. The first line of an input file specifies the
functor $F$ which is written as a term according to the following
grammar:
\begin{equation}\label{eq:functors}
  \begin{aligned}%
    T &::= \texttt{X}
    \mid
    \Pow\, T \mid \Bag\, T \mid \Dist\, T
    \mid \Monval T 
    \mid \Sigma 
    \\
    \Sigma &::= C \mid T + T \mid T \times T \mid T^A
    \quad~
    C ::= \N \mid A
    \quad~
    A ::= \{s_1,\ldots,s_n\} \mid n,
  \end{aligned}
\end{equation}
where $n\in\N$ denotes the set $\{0,\ldots,n-1\}$, the $s_k$ are
strings subject to the usual conventions for variable names
(a letter or an underscore character followed by alphanumeric
characters or underscore),
exponents $F^A$ are written \verb|F^A|,
and $M$ is one of the monoids $(\Z,+,0)$, $(\R,+,0)$,
$(\mathds{C},+,0)$, $(\Pow(64), \cup, \emptyset)$ (the monoid
of $64$-bit words with bitwise $\op{or}$), and $(\N,\max,0)$ (the
additive monoid of the tropical semiring). Note that~$C$ effectively
ranges over at most countable sets, and~$A$ over finite sets. A
term~$T$ determines a functor $F\colon \Set\to\Set$ in the evident
way, with~$\texttt{X}$ interpreted as the argument.

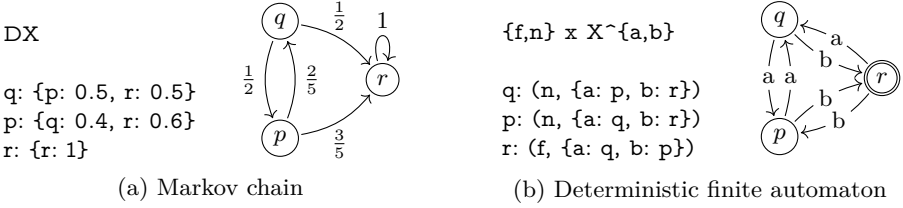
\begin{figure}
  \centering%
  \hfill
  \begin{subfigure}{0.45\textwidth}%
    \begin{minipage}[b]{.55\textwidth}%
\begin{verbatim}
DX

q: {p: 0.5, r: 0.5}
p: {q: 0.4, r: 0.6}
r: {r: 1}
\end{verbatim}
    \end{minipage}
    \hfill%
    \begin{tikzpicture}[coalgebra drawing]
      \node (q1) at (120:9mm) [state] {$q$};
      \node (q2) at (-120:9mm) [state] {$p$};
      \node (q3) at (0:9mm) [state] {$r$};
      \path[->,bend angle=20]
      (q1) edge [left, bend right] node[overlay] {$\frac{1}{2}$} (q2)
      (q1) edge [above,bend left] node[overlay] {$\frac{1}{2}$} (q3)
      (q2) edge [right, bend right] node {$\frac{2}{5}$} (q1)
      (q2) edge [below, bend right,overlay] node {$\frac{3}{5}$} (q3)
      (q3) edge [loop above] node {$1$} (q3);
    \end{tikzpicture}%
    \hfill\hspace*{0pt}%
    \caption{Markov chain}
    \label{subfig:markov}
  \end{subfigure}%
  \hspace{3em}
  \begin{subfigure}{0.43\textwidth}
    \begin{minipage}[b]{.58\textwidth}
\begin{verbatim}
{f,n} x X^{a,b}

q: (n, {a: p, b: r})
p: (n, {a: q, b: r})
r: (f, {a: q, b: p})
\end{verbatim}
    \end{minipage}%
    \hfill%
    \begin{tikzpicture}[coalgebra drawing]
      \node (q) at (120:9mm) [state] {$q$};
      \node (p) at (-120:9mm) [state] {$p$};
      \node (r) at (0:9mm) [state,accepting] {$r$};
      \path[->,bend angle=15,
      every node/.append style={
        shape=circle,
        inner sep=1pt,
        fill=white,
        anchor=center,
      }]
      (q) edge [bend right] node {a} (p)
      (q) edge [bend right] node {b} (r)
      (p) edge [bend right] node {a} (q)
      (p) edge [bend left] node {b} (r)
      (r) edge [bend right] node {a} (q)
      (r) edge [bend left = 20] node {b} (p);
    \end{tikzpicture}
    \caption{Deterministic finite automaton}
    \noshowkeys\label{fig:dfa-picture}
  \end{subfigure}\hfill\,
  \caption{Examples of input files with encoded coalgebras~\cite{wdms21}}
  \vspace{-4mm}
  \label{fig:example-input}
\end{figure}

The remaining lines of an input file specify a finite coalgebra
$c\colon S \to FS$. Each line has the form
$s\texttt{:}\text{\textvisiblespace} t$ for a state $s\in S$, and $t$
represents the element $c(s)\in FS$. The syntax for $t$ depends on the
specified functor~$F$ and follows the structure of the term~$T$
defining~$F$; the details are explained in~\cite[Sec.~3.1.2]{wdms21}.
\autoref{fig:example-input} from \emph{op.~cit.} shows two coalgebras and the
corresponding input files.

After reading the functor term $T$, \copar builds a parser for the
functor-specific input format and then parses the input coalgebra
given in that format into an intermediate format which internally
represents the encoding of the input coalgebra
(\autoref{D:encoding}). For composite functors the parsed coalgebra
then undergoes a substantial amount of preprocessing, which also
affects how transitions are counted; see~\cite[Sec.~3.5]{wdms21} for
more details.

\section{Coalgebraic Partition Refinement}
\label{sec:coalgpartref}

As mentioned in the introduction, the sequential partition refinement
algorithm previously implemented in \copar is based on ideas used
in the Paige-Tarjan algorithm~\cite{PaigeTarjan87} for transition
systems. However, as has been mentioned by Blom and
Orzan~\cite{bo05-01}, the Paige-Tarjan algorithm carefully selects
the block of states to split in each iteration, and the
data structures used for this selection take a lot of memory and
require modification to allow a distributed implementation. Hence,
Blom and Orzan have built their distributed algorithm from a rather simple
sequential partition refinement
algorithm based on what Kanellakis and Smolka refer to as the \emph{naive
  method}~\cite{KanellakisS90}. We now recall this algorithm
and subsequently show how it can be adapted to the coalgebraic level
of generality. 

\subsubsection{Signature Refinement.}
\label{sec:naive-method}

Given a finite labelled transition system with the state set $S$, a
partition on $S$ may be presented by a function
$\pi \colon S \rightarrow \N$, i.e.~two states $s, t \in S$ lie in the
same block of the partition iff $\pi(s) = \pi(t)$.
The \emph{signature} of a state $s \in S$ is the set of outgoing
transitions to blocks of $\pi$:
\begin{equation}\label{eq:sig-lts}
  \sig\pi(s) = \set{(a, \pi(t)) \mid \text{$s \xra{a} t$}} \subseteq \Pow(A \times \N).
\end{equation}
A \emph{signature refinement step} then refines $\pi$ by putting
$s, t \in S$ into different blocks iff
$\sig \pi (s) \ne \sig \pi (t)$. Concretely, we put
$\pi_\text{new}(s) = \hash(\sig \pi (s))$ using a perfect,
deterministic hash function $\hash$. The signature refinement
algorithm~(Fig.~\ref{fig:seq-algo}) starts with a trivial initial
partition on $S$ and repeats the refinement step until the partition
stabilizes,
i.e.~until two subsequent partitions have the same size.

  \begin{figure}
    \begin{algorithm}[H]
      \Variables{old and new partitions represented by
        $\pi, \pi_\text{new} \colon S \rightarrow \N$ with sizes
        $l, l_\text{new}$, resp.; set $H$ for counting block numbers;}
      \ForEach{$s \in S$}{ $\pi_\text{new}(s) \gets 0$ \; }
      $l_{new} \gets 1$\; \While{$l \ne l_\text{new}$}{
        $\pi \gets \pi_\text{new}, H \gets \emptyset$\;
        \ForEach{$s \in S$}{
          $\pi_\text{new}(s) \gets \hash(\sig{\pi}(s))$\;
          $H \gets H \cup \{\pi_\text{new}(s)\}$\; }
        $l \gets l_\text{new}$\; $l_\text{new} \gets |H|$\; }
      \end{algorithm}
      \vspace*{-10pt}
      \caption{Signature refinement for labelled transition systems}
      \label{fig:seq-algo}
      \vspace*{-15pt}
    \end{figure}
    
\subsubsection{Coalgebraic Signature Refinement.}

Regarding a labelled transition system as a coalgebra
$c\colon S \to \Pow(A \times S)$ (\autoref{E:func}\ref{E:func:1}),
signatures are obtained by postcomposing the transition structure with
the partition under the functor:
\begin{align}
  \label{eq:sig-coalg}
  \sig \pi = S \xra{c} \Pow(A \times S) \xra{\Pow(A \times \pi)} \Pow(A \times \N).
\end{align}
The generalisation to coalgebras for arbitrary $F$ is immediate: the
\emph{signature of a state} of an $F$-coalgebra $c\colon S \to FS$
w.r.t.~a partition $\pi$ is given by the function
$\sig \pi = F\pi \cdot c$. %
In the refinement step of the above algorithm two states are
identified by the next partition if they have the same signatures
currently:
\begin{align}
  \label{eq:1}
  \pi_\text{new}(s) = \pi_\text{new}(t)
  \iff
  \sig \pi (s) = \sig \pi (t) \iff
  (F\pi)(c(s)) = (F\pi)(c(t)).
\end{align}
Hence, the algorithm in fact simply applies $F(-) \cdot c$ to the
initial partition corresponding to the trivial quotient
$! \colon S\rightarrow 1$ until stability is reached. Note that this
is precisely the \emph{Final Chain Algorithm} by König and
Küpper~\cite[Alg.~3.2]{KonigK14} computing behavioural equivalence of
a given $F$-coalgebra. Its correctness thus proves correctness of the
\emph{coalgebraic signature refinement} which is the algorithm
in~\autoref{fig:seq-algo} with $\sig \pi = F\pi \cdot c$. Since we
represent functors and their coalgebras by encodings we use an
interface to $F$ to compute signatures based on encodings.
\begin{defn}\label{D:sig-int}
  Given a functor $F$ with encoding $A,\flat_X$, a \emph{signature
    interface} consists of a function
  $\sigop\colon F1 \times \Bag(A \times \N) \to F\N$ such that for
  every finite set~$S$ and every partition $\pi\colon S \to \N$ we
  have
  \begin{equation}\label{eq:sig}
    F\pi = \big(
    FS
    \xrightarrow{\fpair{F!, \flat_S}}
    F1 \times \Bag(A \times S)
    \xrightarrow{F1 \times \Bag(A \times \pi)}
    F1 \times \Bag(A \times \N)
    \xra{\sigop}
    F\N
    \big).
  \end{equation}
\end{defn}
\noindent
Given a coalgebra $c\colon S \to FS$, a state $s\in S$ and a partition
$\pi\colon S \to \N$, the two arguments of $\sigop$ should be
understood as follows. The first argument is the value $F!(c(s)) \in F1$, which intuitively provides
an observable output of the state $s$. The second argument is the bag
$\Bag(A \times \pi)(\flat_S(c(s))$ formed by those pairs
$(a,n)$ of labels $a$ and numbers $n$ of blocks of the partition $\pi$
to which $s$ has an edge; that is, that bag contains one pair $(a,n)$ for each
edge $s \xra{a} s'$ where $\pi(s') = n$. Thus, when supplied with
these inputs, $\sigop$ correctly computes the signature of $s$;
indeed, to see this, precompose
equation~\eqref{eq:sig} with the coalgebra structure~$c$.
\begin{expl}\label{E:sig-int}
  \begin{enumerate}
  \item The constant functor $!C$ has the label set $A=\emptyset$, so
    we have $\Bag(\emptyset \times \N) \cong 1$,
    and we define the function
    $\sigop \colon C \times \Bag(\emptyset \times \N) \rightarrow C$
    by $\sigop(c, *) = c$.
    
  \item\label{E:sig-int:2} The powerset functor $\Pow$ has the label
    set $A=1$, and we define the function
    $\sigop\colon \Pow 1 \times \Bag(1 \times \N) \to \Pow \N$ by
    $\sigop (z, b) = \{ n : b(*,n) \neq 0\}$.
    
  \item The monoid-valued functor $\R^{(-)}$ has the label set
    $A = \R$, and we define the function
    $\sigop \colon \R \times \Bag(\R \times \N) \to
    \R^{(\N)}$ by $\sigop(z, b)(n) = \Sigma\{r \mid b(r, n) \ne 0 \}$.
  \end{enumerate}
\end{expl}
Next we show how signature interfaces can be combined by products
($\times$) and coproducts ($+$). This is the key to the modularity of
the implementation (be it distributed or sequential) of the
coalgebraic signature refinement in \copar.
\begin{constr}\label{C:comb-sig}
  Given a pair of functors $F_1, F_2$ with encodings $A_i,
  \flat_{X,i}$ and signature interfaces $\sigop_{i}$, we put $A = A_1 +
  A_2$ and define the  following functions:
  \begin{enumerate}
  \item for the product functor $F = F_1 \times F_2$ we take
    $\sigop\colon F1 \times \Bag(A\times \N) \to F_1\N \times F_2\N,$ %
    \begin{align*}
      &\sigop(t, b) =
      \big(\sigop_{1}(\pr_1(t),\op{filter}_1(b)),
      \sigop_{2}(\pr_2(t),\op{filter}_2(b))\big).
    \end{align*}
    Here, $\pr_i\colon F1 \to F_i 1$ is the projection map and
    $\op{filter}_i\colon \Bag(A \times \N) \to \Bag(A_i \times \N)$ is
    given by $\op{filter}_i(b)(a,n) = b(\inj_i\,a, n)$,
    where $\inj_i\colon F_i\N \to F\N$ is the injection map.
  \item for the coproduct functor $F = F_1 + F_2$ we take
    \[
      \sigop\colon F1 \times \Bag(A\times \N) \to F_1\N + F_2\N,
      \qquad
      \sigop(\inj_i\,t,b) = \inj_i (\sigop_{i}(t,\op{filter}_i(b))).
    \]
  \end{enumerate}
\end{constr}
\begin{prop}\label{P:comb-sig}
  The functions $\sigop$ defined in \autoref{C:comb-sig} yield
  signature interfaces for the functors $F_1 \times F_2$ and $F_1 +
  F_2$, respectively.
\end{prop}
As a consequence of this result, it suffices to implement signature
interfaces %
only for \emph{basic} functors according to the grammar
in~\eqref{eq:functors}, i.e.~the trivial identity and constant
functors as well as the functors $\Pow$, $\Bag$, $\Dist$ and the
supported monoid-valued functors $M^{(-)}$. Signature interfaces of
products, coproducts and exponents, being a special form
of product, are derived using \autoref{C:comb-sig}. 

Functor composition can be reduced to these constructions by a
technique called \emph{desorting}~\cite[Sec.~8.2]{wdms20},
which transforms a coalgebra of a composite functor into a coalgebra
for a coproduct of basic functors whose signature interfaces can then
be combined by $+$ (see also~\cite[Sec.~3.5]{wdms21}). As for the
previous Paige-Tarjan style algorithm, this leads to the modularity in
the functor of the coalgebraic signature refinement algorithm:
signature interfaces for composed functors are automatically derived
in \copar. Moreover, a new basic functor $F$ may be added by
implementing a signature interface for $F$, effectively extending the
grammar of supported functors in~\eqref{eq:functors} by a clause~$FT$.

\takeout{}%

\takeout{}%

\section{The Distributed Algorithm}
\label{sec:algo}
Our distributed algorithm for coalgebraic signature refinement is a
generalization of Blom and Orzan's original algorithm~\cite{bo05-01} to
coalgebras. We highlight differences to \emph{op.~cit.} at the
end of this section.

\SetKwFunction{workerOf}{workerOf} %
\SetKwFunction{counterOf}{counterOf} %
\SetKwFunction{send}{send}
\newlength\functionLengthSend
\settowidth\functionLengthSend{\send}

\SetKwFunction{distribSum}{distribSum}
\SetKwFunction{sizeOf}{sizeOf}
\SetKwFunction{waitFor}{waitFor}
\SetKwFunction{start}{start}
\SetKwFunction{return}{return}
\SetKw{synchronize}{synchronize}
\SetKwData{DONE}{DONE}
\SetKwData{UPD}{UPD}
\SetKwData{COUNT}{COUNT}
\SetKwProg{onreceive}{on receive}{ do}{end}

We assume a distributed high-bandwidth cluster of $W$ workers
$w_{1}, \ldots, w_{W}$ that is failure-free, i.e.~nodes do not crash,
messages do not get lost and between two nodes the order of messages
is preserved. The communication is based on non-blocking \emph{send}
operations and blocking \emph{receive} operations. Messages are
triples of the form $(\emph{from}, \emph{to}, \emph{data})$, where the
$\emph{data}$ field may be structured and will often
contain a tag to simplify interpretation.

\subsubsection{Description.}

The distributed algorithm is based on the sequential algorithm
presented in~\autoref{fig:seq-algo}, using a distributed hashtable to
keep track of the partition. As for the sequential algorithm, the
input consists of an $F$-coalgebra $(S, c)$ with $|S| = n$ states. We
split the state space evenly among the workers as a preprocessing step.
We write $S_i$ with $|S_i| = n/W$ for the set of states of worker
$w_i$. The input for worker $w_{i}$ is the encoding of that part of
the transition structure of the input coalgebra which is needed to
compute the signatures of the states in~$S_i$. This information is
presented to $w_{i}$ as the list of all outgoing edges of states
of~$S_i$ in the encoding of the coalgebra $(S,c)$, i.e.~the list of
all $s \xra{a} t$ with $s \in S_i$ (cf.~\autoref{D:encoding}). We
refer to the block number $\pi(s)$ of a state $s \in S$ as its \ID.

After processing the input, the algorithm runs in two phases. In the
\emph{Initialization Phase} (\autoref{fig:init}) the workers exchange
update demands about the \ID\/s stored in the distributed hashtable.
If $w_{i}$ has an edge $s \xra{a} s'$ into some state~$s'$ of~$w_{j}$,
then during refinement~$w_{i}$ needs to be kept up to date about the
\ID{} of~$s'$ and thus instructs~$w_{j}$ to do so. Worker $w_{j}$
remembers this information by storing~$w_{i}$ in the set
$\inlistof{s'} = \{w_{i} \mid \exists s \in S_{i}, a \in A.\, s
\xra{a} s'\}$ of incoming edges of~$s'$  (lines~14--16). Hence, for each edge
$s \xra{a} s'$ with $s \in S_{i}$ and $s' \in S_{j}$, worker $w_{i}$
sends a message to~$w_{j}$, informing $w_{j}$ to add $w_{i}$ to
$\inlistof{s'}$ (lines~5--8).
\begin{figure}
  \vspace*{-20pt}
  \begin{algorithm}[H]
      \Variables{Set $V$ of visited states; process count $d$; \\
        for each $s \in S_{i}$ a list $\mathsf{In}_{s}$ of workers with an edge into $s$}
  \end{algorithm}
  \begin{minipage}[c]{0.45\linewidth}
    \begin{algorithm}[H]
      $V \gets \emptyset, d \gets 0$\;
      \ForEach{$s \in S_{i}$}{
        $\inlist \gets []$\;
      }
      \ForEach{edge $s \rightarrow s'$ of $w_{i}$ with $s' \not \in V$}{
        $V \gets V \cup \{s'\}$\;
        \send{$w_{i}, w_{j}, s'$}\;
      }
      \ForEach{$1 \le j \le W$}{
        \send{$w_{i}, w_{j}, \DONE$}\;
      }
      \waitFor{d = W}\;
      \return{$[ \inlist \mid s \in S_{i} ]$}\;
      \setcounter{lastAlgoLine}{\value{AlgoLine}}
    \end{algorithm}
  \end{minipage}
  \begin{minipage}[c]{0.45\linewidth}
    \begin{algorithm}[H]
    \setcounter{AlgoLine}{\thelastAlgoLine}
    \onreceive{$(w_{k}, w_{i}, s)$}{
      $\inlist \gets (w_{k} :: \inlist)$\;
    }

    \vspace{15pt}
    \onreceive{$(\_, \_, \DONE)$}{
      $d \gets d + 1$\;
    }
  \end{algorithm}
\end{minipage}
\caption{Initialization Phase of worker $w_{i}$%
}
\label{fig:init}
\vspace*{-10pt}
\end{figure}

The main phase is the \emph{Refinement Phase} (\autoref{fig:refin}),
mimicking the refinement loop of the undistributed algorithm. In each
iteration all workers compute their part of the new partition,
i.e.~the \ID\/s $h_s = \hash ( \sig{\pi} (s) )$ for each of their
states $s \in S_i$ (line~5). In addition, every worker~$w_{i}$ is responsible
for sending the computed \ID of $s \in S_{i}$ to workers in $\inlist$
that need it for computation of their own signatures in the next
iteration (lines~6--9). The \ID\/s are also sent to a designated worker
\protect\counterOf{$h_{s}$} (lines~10--12). This ensures that \ID\/s are counted
precisely once at the end of the round when the partition size is
computed after all messages have been received (lines~14--17). The actual counting
(line~19) is a primitive operation in the MPI library, for an explicit
$\mathcal{O}(\log W)$ algorithm using messages see e.g.~Blom and
Orzan~\cite[Fig.~6]{bo05-01}. Finally, the workers synchronize before
starting the next iteration (line~20). The refinement phase stops
if two consecutive partitions have the same size (line~2).
\begin{figure}[t]
  \begin{algorithm}[H]
    \Variables{Old, respectively new partitions $\pi, \pi_\text{new}$
      with sizes $l, l_\text{new}$; \\
      finished workers $d$; \ID-counting set $H$;

    }
  \end{algorithm}
  \begin{minipage}[c]{0.53\textwidth}
    \begin{algorithm}[H]
      $\pi_\text{new} \gets 0!, l \gets -1, l_\text{new} \gets 0, H \gets \emptyset$\;
      \While{$l \ne l_\text{new}$}{
        $l \gets l_\text{new}, \pi \gets \pi_\text{new}$\;
        \ForEach{$s \in S_{i}$}{
          $\pi_\text{new}(s) \gets \hash(\sig{\pi}(s))$\;
          \ForEach{$w_{j} \in \inlist$}{
            \send{$w_{i}, w_{j}$,\\
              \hspace*{\functionLengthSend}$ \langle\UPD, s, \pi_\text{new}(s)\rangle$}\;
          }
          \send{$w_{i}$,\\
            \hspace*{\functionLengthSend}\counterOf{$\pi_{new}(s)$},\\
            \hspace*{\functionLengthSend}$\langle\COUNT, \pi_\text{new}(s)\rangle$}\;
        }
        \ForEach{$1 \le j \le W$}{
          \send{$w_{i}, w_{j}, \DONE$}\;
        }
        \waitFor{d = W}\;
        $l \gets l_\text{new}$\;
        $l_\text{new} \gets \distribSum{\sizeOf{H}}$\;
        \synchronize\;
      }
      \setcounter{lastAlgoLine}{\value{AlgoLine}}
    \end{algorithm}
  \end{minipage}
  \begin{minipage}[c]{0.45\textwidth}
    \begin{algorithm}[H]
    \setcounter{AlgoLine}{\thelastAlgoLine}
    \onreceive{$(w_{k}, w_{i}, (\UPD, s, h_{s}))$}{
      $\pi_\text{new}(s) \gets h_{s}$\;
    }

    \vspace{15pt}
    \onreceive{$(w_{k}, w_{i}, (\COUNT, h_{s}))$}{
      $H \gets H \cup \{h_{s}\}$\;
    }

    \vspace{15pt}
    \onreceive{$(\_, w_{i}, \DONE)$}{
      $d \gets d + 1$\;
    }
  \end{algorithm}
\end{minipage}
\vspace*{-10pt}
  \caption{%
    Refinement Phase of worker $w_{i}$%
  }%
  \label{fig:refin}
  \vspace*{-20pt}
\end{figure}
\subsubsection{Correctness.}
\label{sec:correctness}%
\smnote{}
The Initialization Phase (Fig.~\ref{fig:init}) terminates since every
worker reaches line 10, sends \DONE to all workers and thus also
receives it \mbox{(lines~17--19)} a total of $W$ times, allowing it to
progress past line~12. An analogous argument proves termination of
every iteration of the Refinement Phase (Fig.~\ref{fig:refin}). The
sequential algorithm is correct, hence we know the loop of the refinement
phase terminates when all \ID\/s are computed and counted correctly,
since then the distributed and the sequential algorithm compute precisely
the same partitions.

To show that the signatures are computed correctly, we note that if
all \DONE messages have been received in a round, then, by
order-preservation of messages, all messages sent previously in this
round have also been received. This ensures that no workers are
missing from the lists $\inlist$ computed in the Initialization Phase
and that during the Refinement Phase new \ID\/s are sent to all
concerned workers (\autoref{fig:refin}, lines~6--8). This establishes
correctness of the signature computation, and the signatures
coincide on all workers since we assume that the hash function is
deterministic. Finally, the use of the
\counterOf function (line~11) ensures that each \ID{} is included in
the counting set of exactly one worker. Thus, the distributed sum of
the sizes of all counting sets is equal to the size of the
partition.

\subsubsection{Complexity.}
Let us assume that not only states, but also outgoing transitions are
distributed evenly among the workers, i.e.~every worker has about
$m / W$ outgoing transitions. In the Initialization Phase, the loop
sending messages runs in $\CO(\frac m W)$ and receiving takes
$\CO(W \cdot \frac n W)= \CO(n)$, since for worker $w_{i}$ every other
worker $w_{j}$ might have an edge into every state in $S_{i}$. Both
are executed in parallel so in total the phase runs in
$\CO(\max(\frac m W, n)) = \CO(\frac{m} W + n)$.  In the Refinement
Phase, we assume the run time of computing signatures and their hashes is
linear in the number of edges. Then the loop for computing and
hashing~($\CO(\frac m W)$) and counting~($\CO(\frac n W)$) signatures
runs in total in $\CO(\frac{m+n}{W})$, since it is performed by all
workers independently.  Each worker receives at most~$m / W$
\ID-updates each round and the partition size is computable in
$\CO(W)$ giving the complexity of one refinement step in
$\CO(\frac{m+n} W)$. As many as~$n$ iterations %
might be needed for a total complexity of
\(
\CO(\frac m W + n)
+
n \cdot \CO(\frac{n+m} W)
=
\CO\big(\frac{mn +  n^{2}} W + n\big).
\)
\begin{rem}
  The above analysis assumes that signature interfaces are implemented
  with a linear run time in their input bag. This could in fact be
  theoretically realized for all basic functors (whence also for their
  combinations) currently implemented in \copar, which would involve
  using bucket sort for the grouping of bag elements by the
  target block (second component), e.g.~for monoid-valued
  functors. However, since the table used in bucket sort would be
  very large (the size of the last partition) and memory conscience is
  our main motivation, we opted for an implementation using a standard
  $n \log n$ sorting algorithm instead.
\end{rem}

\subsubsection{Implementation details.}

\copar is implemented in Haskell. We were able to reuse, with only
minor adjustments, major parts of the code base of \copar dedicated to
the representation and processing of coalgebras. This includes the
implemented functors and their encodings together with the
corresponding parser and preprocessing algorithms
(see~\autoref{sec:input}). As explained in \autoref{sec:coalgpartref}
the sequential Paige-Tarjan-style algorithm of \copar was not used; we
implemented an additional ``algorithmic frontend'' to our
``coalgebraic backend''. To compute signatures during the Refinement
Phase, each functor implements the signature interface
(\autoref{D:sig-int}), which is written in Haskell as follows:
\begin{lstlisting}
class Hashable (Signature f) => SignatureInterface f where
  type Signature f :: Type
  sig :: F1 f -> [(Label f, Int)] -> Signature f
\end{lstlisting}
We require in the second line a type $\texttt{Signature f}$,
that serves %
as an im\-ple\-men\-ta\-tion-specific
datatype representation of~$F\N$. In the type of $\sigop$, the types
$\mathsf{f}, \mathsf{Label\,f}$ and $\mathsf{F1\,f}$ correspond to
the name of $F$, its label type and the set $F1$, respectively.%
\begin{example}
  The Haskell-implementation of the signature interface for
  the finite power set functor $\Pow$ from
  \autoref{E:sig-int}\ref{E:sig-int:2} is as follows:
\begin{lstlisting}
data P x = P x  -- already defined in CoPaR
type instance Label P = () -- also already defined
instance SignatureInterface P where
  type Signature P = Set Int
  sig :: F1 f -> [((), Int)] -> Set Int
  sig _ = setFromList . map snd
\end{lstlisting}

\end{example}

\noindent
Signature interfaces for the other basic functors according to the grammar
in~\eqref{eq:functors} are implemented similarly. For combined functors \copar automatically
derives their signature interface based on~\autoref{C:comb-sig}.

In the algorithm itself, each worker runs three threads in parallel:
The first thread is for computing, the second one is for sending and
the third one is for receiving signatures. This allows us to keep
calls to the \textsf{MPI} interface separated from (pure) signature
computation, simplifying logic %
and allowing the workers to
scatter the \ID{} of one state while simultaneously computing the
signature of the next one to ensure that neither signature computation
nor network traffic become bottlenecks.  For inter-thread
communication and synchronization we rely on Haskell's \emph{software
  transactional memory}~\cite{hmpj05} to ease concurrent programming,
e.g.~to avoid race conditions.

\subsubsection{Comparison to Blom and Orzan's algorithm.}%
\smnote{}%
We now discuss a few differences of our algorithm to Blom and Orzan's
original one~\cite{bo05-01}.

In Blom and Orzan's algorithm for LTSs the sets $\inlist$ of
$s \in S_{i}$ are in fact \emph{lists} and contain worker $w_{k}$ a
total of $r$ times if there exist $r$ edges from states in~$S_{k}$ to
$s$. This induces a redundancy in messages of \ID{} updates, since
$w_{i}$ sends~$r$ (instead of one) messages with the \ID{} of $s$ to
$w_{k}$. If the LTS has an average fanout of $f$ then each worker has
$t = n/W \cdot f$ outgoing transitions; this is the number of \ID{}
updates received every round. Since there are only $n$ states, at most
$n/t = W/f$ of those messages are necessary. In our scenario, we
have~$W \ll f$ for large coalgebras, hence the overhead becomes
massive; e.g.~for~$W=10, f=100$ already $90\%$ of all \ID{} messages
are redundant.  We use sets instead of lists for $\inlist$ to avoid
this redundancy.  \fbnote[inline]{}

Signature computation and communication do not proceed simultaneously
in Blom and Orzan's original algorithm. However, in their optimized
version~\cite{bo05-06} and in Blom~et~al.'s algorithm for
state labelled continuous-time Markov chains~\cite{bhkp08}
they do.

Another difference of our implementation is that we decided to hash the
signatures directly on the workers of the respective states while Blom
and Orzan decided to first send the signatures to some dedicated
hashing worker who is then (uniquely) responsible for hashing,
i.e.~computing a new \ID. This method allows to compute new \ID\/s in
constant time. However, for more complex functors supported by \copar,
sending signatures could result in very large messages, so we opted
for minimizing network traffic at the cost of slower signature
computation.

\section{Evaluation}
\label{sec:eval}
To illustrate the practical utility and scalability of the algorithm and
its implementation in \copar{}, we report on a number of benchmarks
performed on a selection of randomly generated and real world data. In
previous evaluations of sequential \copar{}~\cite{wdms21}, we were
limited by the 16GB RAM of a standard workstation. Here we demonstrate
that our  distributed implementation fulfills its main objective
of handling larger systems without lifting the memory restriction per
process.
All benchmarks were run on a high performance computing cluster
consisting of nodes with two Xeon 2660v2 ``Ivy Bridge'' chips (10
cores per chip~+ SMT) with 2.2GHz clock rate and 64GB RAM. The nodes
are connected by a fat-tree InfiniBand interconnect fabric with 40
GBit/s bandwidth.
Unless stated otherwise, execution runs were performed using 32 workers on 8 nodes,
resulting in 4 worker processes per node. No process used more than
16GB RAM. Execution times of the sequential algorithm were taken using
one node of the cluster. No times are given for executions that
ran out of 16GB memory previously~\cite{wdms21}; those were not run on the cluster.

\subsubsection{Weighted Tree Automata.} In previous work~\cite{wdms21}, we have
determined the size of the largest weighted tree automata for different
parameters that the sequential version of \copar{} could handle in 16GB
of RAM. Here, we demonstrate that the distributed version can indeed
overcome these memory constraints and process much larger inputs.

Recall from~\autoref{E:func} that weighted tree automata are coalgebras for the
functor $FX = M\times M^{(\Sigma X)}$. For these benchmarks, we use
$\Sigma X = 4\times X^r$ with rank $r\in\{1,\ldots,5\}$ and the monoids
$(2,\vee,0)$ (available as the finite powerset functor in \copar), $(\N, \max, 0)$ and
$(\Pow(64), \cup, \emptyset)$. To generate a random automaton with~$n$ states,
we uniformly chose $k=50\cdot n$ transitions from the set of all possible
transitions (using an efficient sampling algorithm by Vitter~\cite{vitter87})
resulting in a coalgebra encoding with $n'=51\cdot n$ states and
$m = (r+1)\cdot k$ edges. We took care to restrict the state and transition
weights to at most 50 different monoid elements in each example, to avoid the
situation where all states are already distinguished in the first iteration of
the algorithm.

\begin{wrapfigure}{r}{0.5\textwidth}
  \vspace*{-20pt}
  \begin{tikzpicture}%
    \begin{axis}[
      xlabel={Workers used},
      ylabel={\ref{pgfplots:memperworker} Mem./Worker [MB]},
      xmode=log,
      log basis x = {2},
      ymode=log,
      log basis y = {2},
      xmin=8, xmax=128,
      ymin=0, ymax=2600,
      xtick={8, 16, 32, 64, 128},
      ytick={2^7, 2^8, 2^9, 2^10, 2^11},
      legend pos=north east,
      yticklabel style={FAU-Blau},
      ytick style={FAU-Blau},
      ytick pos={left},
      width=5cm
      ]
      
      \addplot[
      color=FAU-Blau,
      mark=square,
      ]
      coordinates {
        (8, 2069)(16,1285)(32,724)(64,392)(128,212)
      };\label{pgfplots:memperworker}
      \legend{}
    \end{axis}
    \begin{axis}[
      xmode=log,
      log basis x = {2},
      xmin=8, xmax=128,
      hide x axis,
      xtick={8, 16, 32, 64, 128},
      ylabel={\ref{pgfplots:comptime} Comp.\ time [s]},
      ymin=0, ymax=220,
      ylabel near ticks, yticklabel pos=right,
      ytick={50, 100, 150, 200, 250},
      legend pos=north east,
      yticklabel style={FAU-RW-Rot},
      ytick style={FAU-RW-Rot},
      ytick pos={right},
      width=5cm
      ]
      
      \addplot[
      color=FAU-RW-Rot,
      mark=x,
      ]
      coordinates {
        (8, 156)(16,99)(32,63)(64,44)(128,52)
      };\label{pgfplots:comptime}
      \legend{}
      
    \end{axis}
  \end{tikzpicture}
  \label{fig:scaling}
  \vspace*{-35pt}
\end{wrapfigure}
\autoref{tab:seq-compare} lists results for both the sequential and
distributed implementation when run on the same input. These are the
largest WTAs for their respective rank and monoid that sequential
\copar{} could handle using at most 16GB of RAM~\cite{wdms21}. In
contrast, the distributed implementation uses less than 1GB per worker
for those examples and is thus able to handle much larger inputs.
Incidentally, the distributed implementation is also faster despite
the overhead incurred by network communication. This can partly be
attributed to the input-parsing stage, which does not need
inter-worker synchronization and is thus perfectly parallelizable.

To test the scaling properties of the distributed algorithm, we ran
\copar{} with the same input WTA but a varying number of worker
processes. For this we chose the WTA for the monoid
$(2, \lor, 0)$ with $\Sigma X = 4 \times X^{5}$ having 86852 states
with 4342600 transitions and file size 186MB.
The figure on the right above
depicts the maximum memory usage per worker
and the overall running time. The results show that both data points
scale nicely with up to 32 workers, but while the running time
even increases when using up to 128 workers, the memory usage
per worker (the main motivation for this work) continues to decrease
significantly. %
\newcommand{\mmc}[1]{\multicolumn{1}{c}{#1}}%
\begin{table}
  \vspace*{-10pt}
\begin{center}
  \begin{tabular}[t]{c@{$\quad$}c@{$\quad$}r@{$\quad$}rr@{$\quad$}r@{$\quad$}c}
    \toprule
    Monoid & $r$ & \mmc{$k$} & \mmc{$n$} &
    \mmc{Mem.~(MB)} & \mmc{Time~(s)} & \mmc{Seq. Time~(s)} \\
    \midrule
    \multirow{6}{*}{$(\mathcal{P}_{\omega}(64), \cup, \emptyset)$}
     & 5 & 4630750   & 92615 & 849 & 61 & 511 \\
     & 4 & 4171550 & 83431   & 663 & 52 & 642 \\
     & 3 & 4721250 & 94425   & 639 & 59 & 528 \\
     & 2 & 6704100 & 134082  & 675 & 76 & 471 \\
     & 1 & 7605350 & 152107  & 642 & 79 & 566 \\
     & 3 & 47212500 & 944250 & 6786 & 675 & -- \\
    \midrule
    \multirow{6}{*}{$(\N, \max, 0)$}
& 5 & 4722550 & 94451  & 871 & 61  & 445 \\
& 4 & 4643950 & 92879  & 754 & 56  & 463 \\
& 3 & 5039950 & 100799 & 628 & 64  & 391 \\
& 2 & 5904200 & 118084 & 633 & 74  & 403 \\
& 1 & 7845650 & 156913 & 677 & 82  & 438 \\
& 3 & 50399500 & 1007990 & 5644 & 645 & -- \\
    \midrule
    \multirow{6}{*}{$(2, \lor, 0)$}
& 5 & 4342600 & 86852  & 701 & 71  & 537 \\
& 4 & 4624550 & 92491  & 728 & 67  & 723 \\
& 3 & 6710350 & 134207 & 825 & 113 & 689 \\
& 2 & 6900000 & 138000 & 715 & 129 & 467 \\
& 1 & 7743150 & 154863 & 621 & 160 & 449 \\
& 3 & 65000000 & 1300000 & 7092 & 1377 & -- \\
\bottomrule
  \end{tabular}
\end{center}
\caption{Maximally manageable WTAs for sequential \copar; ``Mem.'' and
  ``Time'' are the memory and time required for the distributed
  algorithm and are the maximum over all workers. ``Seq.~Time''
  is the time needed by sequential \copar. }%
\fbnote{}
\label{tab:seq-compare}
\vspace*{-20pt}
\end{table}

\subsubsection{PRISM Models.} Finally, we show how our distributed
partition refinement implementation performs on models from the
benchmark suite~\cite{KNP12} of the PRISM model checker~\cite{KNP11}.
These model (aspects of) real-world protocols and are thus a good fit to
evaluate how \copar{} performs on inputs that arise in practice.
Specifically, we use the \emph{fms} and \emph{wlan\_time\_bounded} families of systems.
These are continuous time Markov chains, regarded as coalgebras for
$FX=\R^{(X)}$, and Markov decision processes regarded as coalgebras for
$FX = \N\times \Pow(\N\times (\Dist X))$, respectively. Again, our
translation to coalgebras took care to force a coarse initial partition
in the algorithm.%

The results in \autoref{tab:prism} show that the distributed
implementation is again able to handle larger systems than sequential
\copar{} in 16GB of RAM per process. For the \emph{fms} benchmarks,
the distributed implementation is again faster than the sequential
one. However, this is not the case for the \emph{wlan} examples. The
larger run times might be explained by the much higher number of
iterations of the refinement phase ($i$-column of the table). This
means that only few states are distinguished in each phase, and thus
signatures are re-computed more often and more network traffic is
incurred.
\newcommand{\msp}{\hspace*{2em}}%
\begin{table}[h]
  \begin{center}
    \begin{tabular}[t]{@{$\quad$}l@{$\quad$}r@{$\quad$}rr@{$\quad$}r@{$\quad$}r@{$\quad$}r}
      \toprule
      Model & \mmc{$n$} & \mmc{$m$} & \mmc{Mem.~(MB)} & \mmc{Time~(s)}
      & \mmc{$i$} & \mmc{Seq. Time~(s)} \\
      \midrule
      fms (n=4) & 35910 & 237120 & 13 & 2 & 4 & 4\msp\\
      fms (n=5) & 152712 & 1111482 & 62 & 8 & 5 & 17\msp\\
      fms (n=6) & 537768 & 4205670 & 163 & 26 & 5 & 68\msp\\
      fms (n=7) & 1639440 & 13552968 & 514 & 84 & 5 & 232\msp\\
      fms (n=8) & 4459455 & 38533968 & 1690 & 406 & 7 & --\msp\\
      wlan\_tb (K=0) & 582327 & 771088 & 90 & 297 & 306 & 39\msp\\
      wlan\_tb (K=1) & 1408676 & 1963522 & 147 & 855 & 314 & 105\msp\\
      wlan\_tb (K=2) & 1632799 & 5456481 & 379 & 2960 & 374 & --\msp\\
      \bottomrule
    \end{tabular}
  \end{center}
  \caption{Benchmarks on PRISM models: $n$ and $m$ are the numbers of
    states and edges of the input coalgebra; $i$ is the number of
    refinement steps (iterations). The other columns are analogous to
    \autoref{tab:seq-compare}.}
  \label{tab:prism}
  \vspace*{-20pt}
\end{table}

\section{Conclusions and Future Work}

We have presented a new and simple partition refinement algorithm in
coalgebraic genericity which easily lends itself to a distributed
implementation. Our algorithm is based on König and Küpper's final
chain algorithm~\cite{KonigK14} and Blom and Orzan's signature
refinement algorithm for labelled transition
systems~\cite{bo05-01}. We have provided a distributed implementation
in the tool \copar. Like the previous sequential Paige-Tarjan style
partition refinement algorithm, our new algorithm is modular in the
system type. This is made possible by combining signature interfaces
by product and coproduct, which is used by
\copar for handling combined type functors. Experimentation has shown
that with the distributed algorithm \copar can handle larger state
spaces in general. Run times stay low for weighted tree automata,
whereas we observed severe penalties on some models from the PRISM
benchmark suite. 
An additional optimization of the coalgebraic signature refinement
algorithm should be possible using Blom and Orzan's idea~\cite{bo05-06}
to mark in each iteration those states whose signatures can change in
the next iteration and only recompute signatures for those states in
the next round. This might mitigate the run time penalties we have
seen in some of the PRISM benchmarks.

Further work on \copar concerns symbolic techniques: we have a
prototype sequential implementation of the coalgebraic signature
refinement algorithm where state spaces are represented using BDDs. In
a subsequent step it could be investigated whether this can be
distributed. In another direction the distributed algorithm might be
extended to compute distinguishing formulas, as recently achieved for
the sequential algorithm~\cite{wms21}, for which there is also an
implemented prototype. Finally, there is still work required to
integrate all these new features, i.e.~distribution, distinguishing
formulas, reachability and computation of minimized systems, into one
version of \copar.
\takeout{}%

\subsubsection{Data Availability Statement}
The software \copar{} and the input files that were used to produce the results in this
paper are available for download~\cite{artifact}. The latest version of \copar{} can be
obtained at \url{https://git8.cs.fau.de/software/copar}.

\bibliographystyle{splncs04}
\bibliography{refs}

\vfill

{\small\medskip\noindent{\bf Open Access} This chapter is licensed under the
  terms of the Creative Commons\break Attribution 4.0 International License
  (\url{http://creativecommons.org/licenses/by/4.0/}), which permits use,
  sharing, adaptation, distribution and reproduction in any medium or format, as
  long as you give appropriate credit to the original author(s) and the source,
  provide a link to the Creative Commons license and indicate if changes were
  made.}

{\small \spaceskip .28em plus .1em minus .1em The images or other
third party material in this chapter are included in the\break
chapter's Creative Commons license, unless indicated otherwise in a
credit line to the\break material.~If material is not included in
the chapter's Creative Commons license and\break your intended use
is not permitted by statutory regulation or exceeds the
permitted\break use, you will need to obtain permission directly
from the copyright holder.}

\medskip\noindent\includegraphics{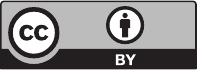}
\clearpage
\appendix
\section{Omitted Details}

First, let us recall from previous work how encodings can be combined
by product and coproduct: 
\begin{notheorembrackets}
\begin{prop}[{\cite[Prop.~3.8]{dmw21}}]\label{P:comb-enc}
  For a pair of functors $F_1, F_2$ with encodings~$A_i, \flat_{X,i}$,
  $i = 1, 2$ we have the following encodings with label set
  $A= A_1 + A_2$:
  \begin{enumerate}
  \item for the product functor $F = F_1 \times F_2$ we take
    \[
      \flat_X\colon
      F_1 X \times F_2 X
      \to
      \Bag((A_1 + A_2) \times X) %
      \qquad
      \flat_X(t)(\inj_i(a), x) = \flat_{X,i}(\pr_i(t))(a,x),
    \]
    where $\inj_i\colon A_i \to A_1+A_2$ and $\pr_i\colon F_1X \times F_2X \to
    F_iX$, $i = 1, 2$, denote the canonical coproduct injections and
    product projections, respectively.
    
  \item for the coproduct functor $F = F_1 + F_2$ we take
    \[
      \flat_X = \big(
      F_1X + F_2 X \xra{\flat_{X,1} + \flat_{X,2}}
      \Bag(A_1 \times X) + \Bag(A_2\times X)
      \xra{\op{can}}
      \Bag((A_1+A_2) \times X)
      \big),
    \]
    where $\op{can} = [\Bag(\inj_1 \times X), \Bag(\inj_2 \times X)]$ is
    the canonical map defined by case distinction on the disjoint
    union. 
  \end{enumerate}
\end{prop}
\end{notheorembrackets}

In the following proof we work with finite products and coproducts 
in lieu of binary ones. Given a finite index set $I$ and a family
$X_i$, $i \in I$, of sets we denote their product and the canonical
projection maps by 
\[
  \prod_{j \in I} X_j
  \xra{\pr_i}
  X_i
  \qquad
  \text{for $i \in I$}.
\]
For every family of maps $f_i\colon X_i \to Y_i$, $i \in I$, we have
the product map
\[
  \prod_{i \in I} X_i
  \xra{\prod_i f_i}
  \prod_{i \in I} Y_i,
  \qquad
  \text{defined by}
  \qquad
  \textstyle\big(\prod_i f_i\big) \big((x_i)_{i\in I}\big) = \big((f_i(x_i))_{i\in I}\big).
\]
The coproduct (disjoint union) of the $X_i$ and the canonical
injection maps are denoted by
\[
  X_i
  \xra{\inj_i}
  \coprod_{j \in I} X_j
  \qquad
  \text{for $i \in I$}.
\]
\begin{remark}\label{R:bags}
  \begin{enumerate}
  \item\label{R:bags:1} Note that for every family of sets $X_i$, $i\in I$, we clearly
    have a canonical isomorphism
    $\Bag (\coprod_i X_i) \cong \prod_i \Bag X_i$ mapping a bag
    $b\colon \coprod_i X_i \to \N$ to the family of bags
    $b \cdot \inj_i\colon X_i \to \N$, $i \in I$. Observe that the
    $i$th component of this isomorphism is a filtering map (cf.~\autoref{C:comb-sig}):
    \[
      \Bag\big(\coprod_{i \in I} X_i\big)
      \xra{\op{filter}_i}
      \Bag X_i,
      \qquad
      \op{filter}_i(b)(x) = b(\inj_i\, x). 
    \]
  \item For the filtering maps from \autoref{C:comb-sig} we clearly have
    \begin{equation}\label{eq:filter}
      \id = \big(
      \Bag(A_i \times \N)
      \xra{\Bag(\inj_i \times \N)}
      \Bag(A \times \N)
      \xra{\op{filter}_i}
      \Bag(A_i \times \N)\big).
    \end{equation}
  \end{enumerate}
\end{remark}

\begin{proof}[\autoref{P:comb-sig}]
  We are going to verify the equation in \autoref{D:sig-int} for the
  product and coproduct functors. First recall from
  \autoref{P:comb-enc} the encodings of these functors. Let $I$ be
  some finite index set and let $F_i$, $i \in I$, be a family of
  functors with encodings $A_i, \flat_{X,i}$, and put
  $A = \coprod_{i\in I} A_i$. 
  \begin{enumerate}
  \item

    For the product functor $F = \prod_{i\in I} F_i$ note first that
    the encoding can be rewritten elementfree as follows: 
    \begin{equation}\label{eq:flat}
      \flat_X = \big(
      \prod_{i\in I} F_iX
      \xra{\prod_i \flat_{X,i}}
      \prod_{i\in I}\Bag(A_i \times X)
      \cong
      \Bag(A \times X)  \big),
    \end{equation}
    where the isomorphism arises from the canonical one in
    \autoref{R:bags}\ref{R:bags:1}.  We now obtain the desired
    equation in \autoref{D:sig-int} by a simple diagram chase:
    \[
      \begin{tikzcd}[column sep = 25, row sep = 30]
        FS = \prod_i F_i S
        \ar[shiftarr = {yshift=22}]{rrr}{\fpair{F!, \flat_S}}
        \ar{rr}{\prod_i \fpair{F_i !, \flat_{S,i}}}
        \ar{d}[swap]{F\pi = \prod_i F_i \pi}
        &
        &
        \prod_iF_i1 \times \Bag(A_i \times S)
        \ar{d}{\prod_i F_i1 \times \Bag(A_i \times \pi)}
        \ar{r}{\cong}
        &
        F1 \times \Bag(A \times S)
        \ar{d}{F1 \times \Bag(A \times \pi)}
        \\
        F\N  = \prod_i F_i \N%
        &
        &
        \prod_iF_i1 \times \Bag(A_i \times \N)
        \ar{ll}[swap]{\prod_i \sigop_i}
        \ar{r}{\cong}
        &
        F1 \times \Bag(A\times \N)
        \ar[shiftarr = {yshift=-21}]{lll}[swap]{\sigop}
        \\
      \end{tikzcd}
    \]
    Note that the horizontal isomorphisms labelled $\cong$ reorder
    factors of the product using the canonical isomorphisms
    $\Bag(A \times X) \cong\prod_i\Bag(A_i \times X)$ for $X = S$ and
    $X = \N$, respectively, making the right-hand square commute due to the
    naturality of the isomorphisms involved. The upper part commutes
    using~\eqref{eq:flat}. Similarly, the lower part is the definition of
    $\sigop$ in elementfree form. The left-hand square commutes by the
    assumption on the $\sigop_i$. Thus, the outside commutes,
    which yields the desired equation.

  \item For the coproduct functor $F = \coprod_i F_i$ we proceed by
    case distinction. More precisely, we verify that the desired
    equation holds when precomposed by every injection map $
    \inj_i\colon F_i S  \to \coprod_i F_iS = FS$. Note first that the $i$th
    coproduct component of the encoding $\flat_X$ is
    \begin{equation}\label{eq:flat-coprod}
      \flat_X \cdot \inj_i = \big(
      F_i X
      \xra{\flat_{X,i}}
      \Bag(A_i\times X)
      \xra{\Bag(\inj_i + X)}
      \Bag(A \times X)\big).
    \end{equation}
    
    Again, we conclude by a simple diagram chase:
    \[
      \begin{tikzcd}[column sep = 35]
        F_iS
        \ar{rd}{\inj_i}
        \ar{dddd}[swap]{F_i \pi}
        \ar{rrr}{\fpair{F_i!,\flat_{S,i}}}
        &&&
        F_i1 \times \Bag(A_i\times S)
        \ar{ld}[swap]{\inj_i \times \Bag(\inj_i \times S)}
        \ar{dddd}{F_i1 \times \Bag(A_i \times\pi)}
        \\
        &
        FS
        \ar{d}[swap]{F\pi}
        \ar{r}{\fpair{F!,\flat_S}}
        &
        F1 \times \Bag(A\times S)
        \ar{d}{F1 \times \Bag(A \times \pi)}
        \\
        &
        F\N
        &
        F1 \times \Bag(A\times \N)
        \ar{l}[swap]{\sigop}
        \\
        &&
        F_i1 \times \Bag(A\times \N)
        \ar{u}[swap]{\inj_i \times \Bag(A\times \N)}
        \ar{d}{F_i1 \times \op{filter}_i}
        \\
        F_i\N
        \ar{ruu}[swap]{\inj_i}
        &&
        F_i1 \times \Bag(A_i\times \N)
        \ar{ll}[swap]{\sigop_i}
        &
        F_i1 \times \Bag(A_i \times \N)
        \ar{lu}[swap,inner sep =0]{F_i1 \times \Bag(\inj_i\times\N)}
        \ar{l}[swap]{\id}
      \end{tikzcd}
    \]
    The desired equation is the commutativity of the rectangle in the
    middle. The upper part commutes by considering the product
    components separately and using~\eqref{eq:flat-coprod} for the
    right-hand one. The left- and right-hand parts clearly
    commutes. The lower part commutes due to the definition of
    $\sigop$, and the lower right-hand triangle commutes
    by~\eqref{eq:filter}. Finally, the outside commutes by the
    assumption on the $\sigop_i$. It follows that the desired
    inner rectangle commutes when precomposed by $\inj_i$, which
    completes the proof.\qed
  \end{enumerate}
\end{proof}%
\end{document}

